\documentclass{ws-procs9x6-cpt19}
\begin{document}

\newcommand{\refeq}[1]{(\ref{#1})}
\def\etal {{\it et al.}}

\def\sla#1{\hbox{{$#1$}\llap{$/$}}}


\title{Renormalization in Nonminimal Lorentz-Violating Field Theory     }

\author{J.R.\ Nascimento,$^1$ A.Yu.\ Petrov,$^{1}$, and C.M.\ Reyes$^2$}

\address{$^1$ Departamento de F\'{\i}sica, Universidade Federal da Para\'{\i}ba,
 \\
Jo\~ao Pessoa, Para\'{\i}ba, Caixa Postal 5008, 58051-970, Brazil}

\address{$^2$ Departamento de Ciencias B\'{a}sicas, Universidad del B\'{\i}o-B\'{\i}o,
  \\
 Chill\'{a}n, Casilla 447, Chile}


\begin{abstract}
We provide the first step towards renormalization in a nonminimal Lorentz-violating model consisting of normal scalars and modified fermions with mass dimension five operators. We compute the radiative corrections corresponding to the scalar self-energy, and we show that some divergencies are improved and in the scalar sector they finite. The pole mass of the scalar two-point function is found and shown to lead to modifications of asymptotic states.
\end{abstract}

\bodymatter
\section{Introduction}
The Standard-Model Extension (SME) can be regarded as an effective 
generalized framework to accommodate possible effects of suppressed
 CPT and Lorentz violation. The SME comprises two different sectors, 
 the minimal sector with operators of mass dimension lower or equal to 
 four~\cite{min}, and the nonminimal sector with higher mass dimensional 
 operators~\cite{nonmin}.  Several bounds have been given using ultrahigh 
 sensitivity experiments which have allowed to test various predictions of Lorentz 
 breakdown in the standard model of particles
and gravity~\cite{datatables}.

A key feature that distinguishes nonminimal models of Lorentz violation
 from minimal ones is the indefinite metric that arises due to the higher 
 time derivative terms. The indefinite metric, as well known, introduces 
 a pseudo unitarity relation for the $S$ matrix, which can imply the loss 
 of conservation of probability. Recently it has been shown that by 
 adopting a Lee-Wick prescription~\cite{Lee-Wick}, it is possible to 
 have a consistent unitary theory~\cite{unitarity}.

 The renormalization of quantum field theories incorporating 
 Lorentz violation can be modified due to radiatively induced operators 
having different structures not initially present in the original 
 Lagrangian~\cite{asymp}. As a general statement, one can 
 say that one-loop Lorentz-violating corrections 
 may have a more dramatic effect on asymptotic states that they do
 have in the typical case. The modified 
 asymptotic space has been mainly shown in computations
  for the mass pole of two-point functions.
 In this work we study the renormalization with particular focus 
 on finite radiative corrections associated to monminimal terms and the effect of 
 modifying the renormalization conditions due to Lorentz violation. 
\section{The Yukawa-like model}
\begin{figure}
\begin{center}
\includegraphics[width=1.5in]{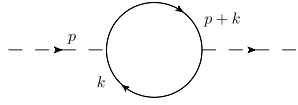}
\end{center}
\caption{Scalar self-energy loop diagram.}
\label{Fig1}
\end{figure}

Consider the Lagrangian density~\cite{ren}
 \begin{eqnarray}
\mathcal L=\frac{1}{2}\partial_{\mu} \phi \partial^{\mu}\phi  
-\frac 1 2M^2\phi^2  
+\bar{\psi}\left(i\sla{\partial}-m -\bar{\alpha} m\sla{n}  \right)\psi +g_2\bar{\psi}
\sla{n}(n\cdot \partial)^2\psi +g\bar{\psi} \phi \psi\,,   \nonumber \\
\end{eqnarray}
where $n^{\mu}$ is a preferred four-vector, $g_2$ is a constant with dimension of $(\text mass)^{-1}$ 
and $g$ is a typical Yukawa coupling.
As a first example of quantum corrections in our model, 
we study the contribution with two external scalar legs depicted at
Fig.~\ref{Fig1}. 

It is represented by the integral
\begin{eqnarray}
i\Pi(p)&=&-\frac{g^2}{2}\phi(-p)\phi(p)
\int\frac{d^4k}{(2\pi)^4}
  \frac{ {\rm Tr}  \left(  \left(  Q_{\mu} \gamma^{\mu}+m \right)
(R_{\nu} \gamma^{\nu}+m)\right)}{(Q^2-m^2)(R^2-m^2)}\,,  
\end{eqnarray}
where we define
\begin{eqnarray}\label{QR}
Q_{\mu}(k)&=&k_{\mu}-\bar{\alpha} m n_{\mu}-g_2n_{\mu}(n\cdot k)^2\,, \\ 
R_{\mu}(p+k)&=&p_{\mu}+k_{\mu}-\bar{\alpha} m n_{\mu}-g_2n_{\mu}
(n\cdot (p+k))^2\,.
\end{eqnarray}
We can write the correction to the scalar propagator up to second-order in $p$ 
as
\begin{eqnarray}\label{Pi}
\widetilde \Pi (p)&=&m^2 q_0+p^2q_1+(n\cdot p)^2q_n\,,
\end{eqnarray}
where
\begin{eqnarray}\label{q_0}
q_0&=&-\frac{i}{48\pi^2g_2^2m^2}
+\frac{i}{16\pi^2}
\left(2(3\gamma_E-1)-0.46    \right.  \\   && \left. -3\ln \left(\frac{g_2^2m^2}{4}\right)\right)  \nonumber \,,
\\ \label{q_1}
q_1&=&-\frac{ i}{12\pi^2}\left(-5+6\gamma_E  -0.46-3\ln\left(\frac{g_2^2m^2}{4}\right)     \right)  \,,
\\ \label{q_n}
q_n&=&\frac{i}{\pi^2}\,.
\end{eqnarray}
The renormalized scalar two-point function is given by
\begin{eqnarray}\label{renPi}
(\Gamma_{\phi,R}^{(2)})^{-1}=p^2-M^2+\Pi_R(p)\,,
\end{eqnarray}
with
\begin{eqnarray}
\Pi_R(p)=p^2A_{\phi} +m^2 B_{\phi}+(n\cdot p)^2C_{\phi}\,,
\end{eqnarray}
where $A_{\phi} $, $B_{\phi}$ and $C_{\phi}$ are constants that can be deduced from 
the expressions~(\ref{q_0}),~(\ref{q_1}),~(\ref{q_n}) being
\begin{eqnarray}
iA_{\phi}&=&-2g^2q_1  \,,  \nonumber \\ 
iB_{\phi}&=&-2g^2q_0 \,,\nonumber\\
 iC_{\phi}&=&-2g^2q_n\,.
\end{eqnarray}
We consider the ansatz $\bar P^2_{\phi}= p^2- { M}^2_{{\rm ph}}+\bar y(n\cdot p)^2$
in terms of the two unknown constants $M_{{\rm ph}}$ and $\bar y$,
Both constants can be determined
using the renormalization condition
\begin{eqnarray}\label{scalarcond}
(\Gamma_{\phi,R}^{(2)})^{-1}|_{\bar P_{\phi}^2=0}=0\,.
\end{eqnarray}
Hence, from (\ref{renPi}) replacing the value of $p^2$ 
and using the condition (\ref{scalarcond}), we arrive at the equation
\begin{eqnarray}
0&=&{ M}^2_{{\rm ph}}   -\bar y(n\cdot p)^2-M^2+  A_{\phi} 
\left({ M}^2_{{\rm ph}}   -\bar y(n\cdot p)^2\right) +B_{\phi}m^2+C_{\phi}(n\cdot p)^2\,. \nonumber \\
\end{eqnarray}
Due to the independence of each term, we find the two constants
\begin{eqnarray}
{ M}^2_{{\rm ph}} =\frac{(M^2-B_{\phi}m^2)}{(1+A_{\phi})}\,, \qquad \bar y =\frac{C_{\phi}}{1+A_{\phi}}\,,
\end{eqnarray}
and in consequence also the scalar pole mass $\bar P_{\phi}^2$ which 
dictates how asymptotic states propagate. 
Substituting the above expressions in Eq.~(\ref{renPi}) and using the normalization condition
\begin{eqnarray}
\frac{d(\Gamma_{\phi,R}^{(2)})^{-1}}{d\bar P^2_{\phi}}|_{\bar P^2_{\phi}=0}=Z_{\phi}^{-1}\,,
\end{eqnarray}
we identify 
the finite wave function renormalization constant $Z_{\phi}^{-1}=1+A_{\phi}$.
\section{Conclusions}
For nonminimal Lorentz-violating models, one should, in general, expect an indefinite metric leading to a nontrivial structure of poles and a pseudo unitary relation for the $S$ matrix. Also, the nonstandard structure of radiative corrections in general 
leads to a modified asymptotic space.
For the model we have focus on we have considered 
 the prescription for locating the poles to be dictated by unitarity conservation requirements. We have found that some radiative corrections are improved, and in fact  they are finite in the scalar sector. The pole extraction for the Yukawa-like model has been successfully provided
 in the scalar sector.
\section*{Acknowledgments}
The work 
by A. Yu. P. has been supported by the
CNPq project No. 303783/2015-0. CMR acknowledges support by FONDECYT grant 1191553.

\end{document}